\documentclass[preprint,prl,showpacs,nofootinbib]{revtex4}
\usepackage[mathscr]{euscript}
\usepackage{graphicx}
\usepackage{float,epsfig}
\usepackage{dcolumn}
\usepackage{bm}
\usepackage{graphicx}
\usepackage{dcolumn}
\usepackage{amsmath,amssymb,amsthm}
\usepackage[colorlinks=true,linkcolor=blue]{hyperref}
\usepackage{booktabs}

\newcommand{\bea}{\begin{eqnarray}}
\newcommand{\eea}{\end{eqnarray}}
\newcommand{\beq}{\begin{equation}}
\newcommand{\eeq}{\end{equation}}
\newcommand{\nn}{\nonumber}
\def\/{\over}

\begin{document}
\title{Detecting modified vacuum fluctuations due to presence of a boundary by means of the geometric phase}
\author{   Hongwei Yu$^{1,2 }$
 and Jiawei Hu$^{2}$  }
\affiliation{$^1$ Center for Nonlinear Science and Department of Physics, Ningbo
University, Ningbo, Zhejiang 315211, China\\
$^2$ Department of Physics and Key Laboratory of Low
Dimensional Quantum Structures and Quantum
Control of Ministry of Education,\\
Hunan Normal University, Changsha, Hunan 410081, China}


\begin{abstract}

We study the geometric phase acquired by an inertial atom whose
trajectories are parallel to a reflecting boundary due its
coupling to vacuum fluctuations of  electromagnetic fields, by treating
the atom as an open quantum system in a bath of the fluctuating vacuum
fields, and show that the phase is  position dependent as a
result of  the presence of the boundary which modifies  the field quantum
fluctuations. Our result therefore suggests a possible way of detecting  vacuum fluctuations in experiments
involving geometric phase.

\end{abstract}

\maketitle

Quantum theory has profoundly changed our conception of vacuum as a
synonym of nothingness.  As an inevitable consequence necessitated by  the
uncertainty principle, vacuum fluctuates and thus may have rich
structures. An intriguing issue is what are the physical
consequences of vacuum fluctuations that exist all the time and whether these fluctuations can be directly detected. In this
regard, let us note that the effects of vacuum fluctuations  in free
space may not be always observable, let alone a direct observation, since some physical quantities,
energy for instance, are not well-defined in vacuum and one has to
invoke  certain renormalization schemes to make them finite.
However, changes in the vacuum fluctuations, e.g., those caused by
the presence of boundaries, usually exhibit normal behaviors and can
produce observable effects. The Lamb shift~\cite{lamb} and  the
Casimir~\cite{casimir1} (and the Casimir-Polder~\cite{casimir2,
casimir3}) effects are the two most well-known examples, which have been
precisely measured. Other examples of novel effects that arise as a
result of the vacuum fluctuations include but are by no means limited to
the light-cone fluctuations when gravity is
quantized~\cite{lightcone}, the Brownian motion of test particles in
an electromagnetic vacuum~\cite{brown}, and modifications of
radiative properties of atoms in cavities such as the natural
lifetimes and energy level shifts which have been demonstrated in
experiments~\cite{cavity}. In this paper, we show that the vacuum
fluctuations may also be directly detected though the measurement of
geometric phase. As a related issue of vacuum fluctuation detection, it is worth noting that the quantum vacuum fluctuation in the position of
a mechanical system has recently been clearly detected using a nanomechanical resonator~\cite{AHS}.

The geometric phase is an important concept in quantum theory. In
1984, Berry studied the dynamics of a closed quantum system whose
Hamiltonian varies  adiabatically in a cyclic way, and found that there is,
besides the familiar dynamical phase, an additional
 phase due to the geometry of the path enclosed during the unitary evolution
of the system in the parameter space~\cite{berry}. Ever since the
inception, the geometric phase has aroused broad interest and has
been extensively studied, both theoretically and
experimentally~\cite{GPbook}. Recent concerns about the geometric
phase mainly focus on its potential of performing fault-tolerant
quantum computation~\cite{fault-tolerant}.  Because of the
inevitable interactions between the qubits and the environment, a
pure state will generically be  driven to a mixed state. To deal with the effect of the environment, many attempts have been made which generalize the geometric phase of a closed system undergoing a unitary evolution to an open quantum system which undergoes a nonunitary one~\cite{Uhlmann, Sjoqvist, Singh, Tong, Wang}. Remarkably, experiments have demonstrated both the geometric phase of a mixed state  undergoing a cyclic unitary evolution ~\cite{Du, Marie} and that of an open system undergoing nonunitary one~\cite{critical}.

In fact, the impact of environment on the geometric phase
is a crucial  issue in any practical implementations of quantum
computing. The effects of different kinds of decoherence sources on
the geometric phase have been analyzed~\cite{Carollo, Rezakhani,
Lombardo, chen, Marzlin}. It is remarkable that when the temperature
is absolute zero, there is still a correction to the geometric phase
caused by the environment, i.e., a reservoir of  vacuum fluctuations~\cite{Lombardo, chen, Marzlin}. However, this kind of inevitable vacuum
fluctuation induced geometric phase  is in general unobservable,
since any phase variation is observed usually via some
kind of interferometry between the involved state and certain
selected reference states which are both inseparably and equally coupled to
vacuum. Nevertheless, if, somehow,
 vacuum fluctuations are modified, then the geometric phase of the
nonunitary evolution of an open system caused by its coupling to
vacuum may become potentially observable. Here, we show that the modification of vacuum
fluctuations induced by  the presence of boundaries provides  such a possibility to unveiling quantum vacuum fluctuations via geometric phase.
At this point, it is worth noting that geometric phase has recently been proposed as a possible way to detect the Unruh effect in Ref.~\cite{Martin} and later in Ref.~\cite{HuYu}.


The system we study contains an inertial two-level atom in
interaction with a bath of fluctuating quantum electromagnetic
fields in vacuum at a fixed  distance to a reflecting boundary. The
Hamiltonian of the whole system takes the form $H=H_s+H_\phi+H'.$
Here $H_s$ is the Hamiltonian of the atom, which, for simplicity, is
taken to be $H_s={1\over 2}\,\hbar\omega_0\sigma_3,$ where
$\sigma_3$ is the Pauli matrix, and $\omega_0$ is the energy-level
spacing of the atom. $H_\phi$ is the Hamiltonian of the
electromagnetic field, of which the explicit form is not relevant
here. In the multipolar coupling scheme~\cite{CPP95}, the interaction Hamiltonian $H'$ takes the
form $H'(\tau)=-e\,\textbf{r} \cdot
\textbf{E}(x(\tau))=-e\sum_{mn}\textbf{r}_{mn}\cdot
\textbf{E}(x(\tau))\sigma_{mn},$ where {\it e} is the electron
electric charge, $e\,\bf r$ the atomic electric dipole moment, and ${\bf E}(x)$ the electric-field strength. Here, the dipole moment must be kept fixed with respect to the proper frame of reference of the atom; otherwise, the rotation of the dipole moment will bring in extra time dependence in addition to the intrinsic time evolution~\cite{Takagi}. Since neither $\textbf{r}$ nor $\textbf{E}(x)$ is a world vector, the interaction Hamiltonian $H'$ is ambiguous when we deal with the situation of moving atoms. However, we can write the interaction Hamiltonian $H'$ in a coordinate invariant form as $H'=-e\,r^{\mu}F_{\mu\nu}(x)\,u^{\nu}$, where $F_{\mu\nu}$ is the field strength, $r^{\mu}$ is a four-vector such that its temporal component in the frame of the atom vanishes and its spatial components in the same frame are given by $\textbf{r}$, and $u^{\nu}$ is the four velocity of the atom. Since we choose to work in the frame of the atom, $u^{\nu}=(c,0,0,0)$, and this coordinate invariant interaction Hamiltonian reduces to the form given above~\cite{Takagi,Borde}. The dynamical evolution of the two-level-atom subsystem will be studied in the paradigm of open quantum systems. Here, let us note that the theory of open quantum systems has been fruitfully applied to understand, from a  perspective different from the traditional,  the Unruh, Hawking and Gibbons-Hawking effects, in Refs.~\cite{Benatti1}, \cite{yu3}, and~\cite{yu4}, respectively.

The initial state of the whole system is characterized by the total
density matrix $\rho_{tot}=\rho(0) \otimes |0\rangle\langle0|$, in
which $\rho(0)$ is the initial reduced density matrix of the atom,
and $|0\rangle$ is the vacuum state of the field. In the frame of
the atom, the evolution in the proper time $\tau$ of the total
density matrix $\rho_{tot}$ satisfies
\begin{equation}
\frac{\partial\rho_{tot}(\tau)}{\partial\tau}=-{i\/\hbar}[H,\rho_{tot}(\tau)]\;.
\end{equation}
We assume that the interaction between the atom and the field is
weak. In the limit of weak coupling, the evolution of the reduced
density matrix $\rho(\tau)$ can be written in the
Kossakowski-Lindblad form~\cite{Lindblad, Benatti1, Benatti2, pr5}
\begin{equation}\label{master}
{\partial\rho(\tau)\over \partial \tau}= -{i\/\hbar}\big[H_{\rm eff},\,
\rho(\tau)\big]
 + {\cal L}[\rho(\tau)]\ ,
\end{equation}
where
\begin{equation}
{\cal L}[\rho]={1\over2} \sum_{i,j=1}^3
a_{ij}\big[2\,\sigma_j\rho\,\sigma_i-\sigma_i\sigma_j\, \rho
-\rho\,\sigma_i\sigma_j\big]\ .
\end{equation}
The matrix $a_{ij}$ and the effective Hamiltonian $H_{\rm eff}$ are determined by the Fourier and Hilbert transforms of the correlation functions,
\begin{equation}
G^{+}(x-y)={e^2\/\hbar^2} \sum_{i,j=1}^3\langle -|r_i|+\rangle\langle +|r_j|-\rangle\,\langle0|E_i(x)E_j(y)|0 \rangle\;,
\end{equation}
which are defined as follows:
\begin{equation}
{\cal G}(\lambda)=\int_{-\infty}^{\infty} d\tau \,
e^{i{\lambda}\tau}\, G^+\big(x(\tau)\big)\; ,
\end{equation}
\begin{equation}
{\cal K}(\lambda)= \frac{P}{\pi i}\int_{-\infty}^{\infty} d\omega\
\frac{ {\cal G}(\omega) }{\omega-\lambda} \;.
\end{equation}
Then the coefficients of the Kossakowski matrix $a_{ij}$ can be expressed as
\begin{equation}
a_{ij}=A\delta_{ij}-iB
\epsilon_{ijk}\delta_{k3}-A\delta_{i3}\delta_{j3}\;,
\end{equation}
in which
\begin{equation}\label{abc}
A=\frac{1}{4}[{\cal G}(\omega_0)+{\cal G}(-\omega_0)]\;,\;~~~
B=\frac{1}{4}[{\cal G}(\omega_0)-{\cal G}(-\omega_0)]\;.
\end{equation}
The effective Hamiltonian $H_{\rm eff}$ contains a correction term,
the so-called Lamb shift, and one can show that it is given
by replacing $\omega_0$  in $H_s$ with a renormalized energy-level
spacing $\Omega$ as follows~\cite{Benatti1}:
\begin{equation}\label{heff}
H_{\rm eff}=\frac{1}{2}\hbar\Omega\sigma_3={\hbar\over 2}\{\omega_0+{i\/2}[{\cal
K}(-\omega_0)-{\cal K}(\omega_0)]\}\,\sigma_3\;.
\end{equation}

 Assuming the initial state of the atom is
$|\psi(0)\rangle=\cos{\theta\/2}|+\rangle+\sin{\theta\/2}|-\rangle$, one can show that the time-dependent reduced density matrix of the atom is given by
\begin{equation}\label{dens}
\rho(\tau)=\left(
\begin{array}{ccc}
e^{-4A\tau}\cos^2{\theta\/2}+{B-A\/2A}(e^{-4A\tau}-1) & {1\/2}e^{-2A\tau-i\Omega\tau}\sin\theta\\ {1\/2}e^{-2A\tau+i\Omega\tau}\sin\theta & 1-e^{-4A\tau}\cos^2{\theta\/2}-{B-A\/2A}(e^{-4A\tau}-1)
\end{array}\right)\;,
\end{equation}
which evolves nonunitarily. The geometric phase for a mixed state under a nonunitary evolution
can be defined as~\cite{Tong}
 \beq\label{gp} \gamma_g=\arg \left(
\sum\limits_{k=1}^N \sqrt{\lambda_k(0)\lambda_k(T)}\langle
\phi_k(0)|\phi_k(T)\rangle e^{-\int_0^T \langle \phi_k(\tau)|\dot
\phi_k(\tau) \rangle d\tau} \right)\;, \eeq
where $\lambda_k(\tau)$ and $|\phi_k(\tau)\rangle$ are the eigenvalues and eigenvectors of
the reduced density matrix $\rho(\tau)$. In order to find the
geometric phase,  we first calculate the eigenvalues of the density
matrix (\ref{dens}) to get $
\lambda_\pm(\tau)={1\/2}(1\pm\eta)\;,$ in which
$\eta=\sqrt{\rho_3^2+e^{-4A\tau}\sin^2\theta}$ and
$\rho_3=e^{-4A\tau}\cos\theta+{B\over A}(e^{-4A\tau}-1)$. It is easy to see
that $\lambda_-(0)=0$. As a result,  contribution only
comes from the eigenvector corresponding to $\lambda_+$ ,
\beq
|\phi_+(\tau)\rangle=\sin{\theta_{\tau}\/2}|+\rangle+\cos{\theta_{\tau}\/2}e^{i\Omega\tau}|-\rangle\;,
\eeq where \beq\label{tan}
\tan{\theta_{\tau}\/2}=\sqrt{\eta+\rho_3\/\eta-\rho_3}\;. \eeq
The
geometric phase can then be calculated directly using Eq.~(\ref{gp}),
\beq\label{gp1}
\gamma_g=-\Omega\int_0^T\cos^2{\theta_{\tau}\/2}\,d\tau\;. \eeq

 Now, we  calculate the geometric phase of an atom in the vicinity of a reflecting boundary. To do so, we need the two point functions for the electric fields,
 which can be found from those of  the four-potentials,
 \beq
D^{\mu\nu}(x,x')=\langle0|A^\mu(x)A^\nu(x')|0\rangle=D_0^{\mu\nu}(x,x')+D_b^{\mu\nu}(x,x')\;,
\eeq
in which $D_0^{\mu\nu}(x,x')$ is the two point function in the
Minkowski vacuum without boundaries, and $D_b^{\mu\nu}(x,x')$ is the correction
induced by the presence of the boundary which can be calculated using the method of images. In the Feynman gauge,  we have, at a distance $z$ from the boundary,
\begin{eqnarray}
&&D_0^{\mu\nu}(x,x')={\hbar\/4\pi^2\varepsilon_0c}{\eta^{\mu\nu}\/{
[(ct-ct^\prime-i\varepsilon)^2-(x-x^\prime)^2-(y-y^\prime)^2-(z-z^\prime)^2]}}\;,\\
&&D_b^{\mu\nu}(x,x')=-{\hbar\/4\pi^2\varepsilon_0c}{{\eta^{\mu\nu}+2n^\mu n^\nu}\/{
[(ct-ct^\prime-i\varepsilon)^2-(x-x^\prime)^2-(y-y^\prime)^2-(z+z^\prime)^2]}}\;,
\end{eqnarray}
where $\eta^{\mu\nu}=diag(1,-1,-1,-1)$, $n^\mu=(0,0,0,1)$, and
$\varepsilon\rightarrow+0$. The electric field two-point functions
can be expressed as a sum of  the Minkowski vacuum term and a
correction term due to the boundary:
\begin{eqnarray}\label{2p1}
\langle E_i(x(\tau))E_j(x(\tau'))\rangle=\langle
E_i(x(\tau))E_j(x(\tau'))\rangle_0 +\langle
E_i(x(\tau))E_j(x(\tau'))\rangle_b\;,
\end{eqnarray}
where
\begin{eqnarray}\label{2p2}
\langle0|E_i(x(\tau))E_j(x(\tau'))|0\rangle_0&=&{\hbar c\/4\pi^2\varepsilon_0}(\partial
_0\partial_0^\prime\delta_{ij}-\partial_i\partial_j^\prime)\nonumber\\&&\times
{1\/(x-x')^2+(y-y')^2+(z-z')^2-(ct-ct'-i\varepsilon)^2}\;,\label{ee1}
\end{eqnarray}
\begin{eqnarray}\label{2p3}
\langle0|
E_i(x(\tau))E_j(x(\tau'))|0\rangle_b&=&-{\hbar c\/4\pi^2\varepsilon_0}[\,(\delta_{ij}-2n_in_j)\,\partial
_0\partial_0^\prime-\partial_i\partial_j^\prime\,]\nonumber\\&&\times
{1\/(x-x^\prime)^2+(y-y^\prime)^2+(z+z^\prime)^2-(ct-ct^\prime-i\varepsilon)^2}\;.\label{ee2}
\end{eqnarray}
Here $\partial^\prime$ denotes the
differentiation with respect to $x^\prime$.

Let us now consider an atom moving in the $x$-direction with a
constant velocity $v$ at a distance $z$ from the plane, so the
trajectory is given by
\begin{eqnarray}\label{traj}
t(\tau)=\gamma\tau\;, \ \ \ x(\tau)=x_0+v\gamma\tau\;, \ \ \
y(\tau)=y_0\;,\ \ \ z(\tau)=z\;,
\end{eqnarray}
where $\gamma=(1-v^2/c^2)^{-{1\/2}}$. Here let us note that the electric-field two-point functions  Eqs.~(\ref{2p1})-(\ref{2p3}), and the trajectory Eq.~(\ref{traj}) are described in the laboratory frame. Since the evolution of the atom is studied in the frame of the atom, a Lorentz transformation is required to get the electric field two-point functions in the proper frame of the atom from Eq.~(\ref{2p1}) to Eq.~(\ref{2p3}):
\bea
&&\langle0|E_i(x(\tau))E_j(x(\tau'))|0\rangle_0
={\hbar c\/\pi^2\varepsilon_0}{\delta_{ij}\/(c\,\Delta\tau-i\varepsilon)^4}\;,\nonumber\\
&&\langle0|E_x(x(\tau))E_x(x(\tau'))|0\rangle_b=\langle0|E_y(x(\tau))E_y(x(\tau'))|0\rangle_b
=-{\hbar c\/\pi^2\varepsilon_0}{c^2\Delta\tau^2+4z^2\/[\;(c\,\Delta\tau-i\varepsilon)^2-4z^2]^3}\;,\nonumber\\
&&\langle0|E_z(x(\tau))E_z(x(\tau'))|0\rangle_b
={\hbar c\/\pi^2\varepsilon_0}{1\/[\;(c\,\Delta\tau-i\varepsilon)^2-4z^2]^2}\;.
\eea
The correlation function and its Fourier transform can be calculated as follows:
\beq
G^+(x)=\sum_i{e^2\/\hbar^2}\,|\langle -|r_i|+\rangle|^2\langle0|E_i(x)E_i(y)|0 \rangle\;,
\eeq
\beq\label{fourier}
{\cal G}(\lambda)=\sum_i{e^2|\langle -|r_i|+\rangle|^2\lambda^3\/3\pi\varepsilon_0\hbar c^3}(1-f_i(\lambda,z))\theta(\lambda)\;,
\eeq
where
\beq\label{fourier1}
f_x(\lambda,z)=f_y(\lambda,z)={3c^3\/16\lambda^3z^3}\bigg[{2\lambda z\/c}\cos{2\lambda z\/c}+({4\lambda^2z^2\/c^2}-1)\sin{2\lambda z\/c}\bigg]\;,
\eeq
\beq\label{fourier2}
f_z(\lambda,z)={3c^3\/8\lambda^3z^3}\bigg[{2\lambda z\/c}\cos{2\lambda z\/c}-\sin{2\lambda z\/c}\bigg]\;,
\eeq
and $\theta(\lambda)$ is the standard step function. Thus the coefficients of the Kossakowski matrix $a_{ij}$ and the effective level spacing of the atom can be written as
\beq
A=B={\gamma_0\/4}\sum_i\alpha_i(1-f_i(\omega_0,z))\;,
\eeq
\beq\label{lm}
\Omega=\omega_0+{\gamma_0\/2\pi\omega_0^3}\sum_i\,P\int_0^\infty
d\omega\,\omega^3\alpha_i\big(1-f_i(\omega_0,z)\big)\bigg({1\/\omega+\omega_0}-{1\/\omega-\omega_0}\bigg)\,\;, \eeq
where $\gamma_0=e^2|\langle -|{\bf
r}|+\rangle|^2\,\omega_0^3/3\pi\varepsilon_0\hbar c^3$ is the spontaneous emission rate in vacuum without boundaries, and $\alpha_i=|\langle
-|r_i|+\rangle|^2/|\langle -|{\bf r}|+\rangle|^2\,.$ Then the
geometric phase can be found  using Eq.~(\ref{gp1}),
\beq\label{bd}
\gamma_{g}=-\int_0^{T}{1\over2}\bigg(1-\frac{1-e^{4A\tau}+\cos\theta}{\sqrt{e^{4A\tau}\sin^2\theta
+(1-e^{4A\tau}+\cos\theta)^2}}\bigg)\,\Omega\,d\tau\;. \eeq So, the
phase accumulates as the system evolves, although the accumulation
with time is not linear as in the unitary evolution case. For a
single period of evolution, the  result of this integral
can be analytically expressed as \beq
\gamma_{g}={\Omega\/\omega_0}\big(F(2\pi)-F(0)\big)\;,
\eeq
where function $F(\varphi)$ is defined as
\bea
F(\varphi)&=&-{1\/2}\,\varphi-{1\/8A}\ln \left({-{1\/2}Q^2+e^{4A\varphi/\omega_0}}+S(\varphi) \right)\nn\\
&&-{1\/8A}\text{sgn}(Q)\ln \left({ {2Q^2e^{-4A\varphi/\omega_0}-Q^2 }+2Q\,S(\varphi)\,{e^{-4A\varphi/\omega_0)}}} \right)\;,
\eea
in which $S(\varphi)=\sqrt {e^{8A\varphi/\omega_0}-{e^{4A\varphi/\omega_0}}Q^2+Q^2}\;$, $Q=1+\cos\theta$, and $\text{sgn}(Q)$ is the standard sign function.

In order to examine the behaviors of this phase, we perform, for
small $\gamma_0/\omega_0$, which is true in our current discussions
as we will see later, a series expansion of the geometric phase for
a single quasi-cycle and find, to the first order,~\footnote{Here we
have omitted the Lamb shift terms, since it is obvious that these
terms contain a factor $\gamma_0/\omega_0$ and they will only
contribute to the phase at the second and higher orders of
$\gamma_0/\omega_0$.}
\beq\label{GPb}
\gamma_{g}\approx-\pi(1-\cos\theta)-\pi^2{\gamma_0\/2\omega_0}\sum_i\alpha_i
(1-f_i(\omega_0,z))(2+\cos\theta)\sin^2\theta\;.
\eeq
The first term
$-\pi(1-\cos\theta)$ in the above equation is what we would have
obtained if the system were isolated from the environment, i.e., a
bath of fluctuating vacuum electromagnetic fields and the second
term is the correction induced by the interaction between the atom
and the environment. Here $f_i(\omega_0,z)$ are oscillating
functions of distance $z$ with a position-dependent amplitude. For
an atom polarized in an arbitrary direction, the polarizations of
the atom in the tangential directions and in the normal direction of
the boundary contribute differently to the correction of the
geometric phase. If the atom is polarized in the tangential
direction, as the atom approaches the boundary $(z\rightarrow0)$,
the correction of the geometric phase vanishes, since
$f_x(\omega_0,z)$ and $f_y(\omega_0,z)$ approach zero, which can be
attributed to the fact that the tangential components of the
electric field vanish on the conducting plane. However, if the atom is
polarized in the normal direction, $f_z(\omega_0,z)\rightarrow-1$ as
$z\rightarrow0$, and the correction of the geometric phase is twice
that of the free space case. This can be understood as the fact that
the reflection at the boundary doubles the normal component of the
fluctuating electric field. When the distance $z$ approaches
infinity, the modulation functions $f_i(\omega_0,z)$ approach zero,
and the result reduces to that of the unbounded Minkowski vacuum case. So,
due to the modification of the vacuum fluctuations caused by the
reflecting plane, the vacuum fluctuation induced geometric phase becomes
position dependent. Now let us estimate how large the phase
difference is. If we assume that $|\langle -|{\bf r}|+\rangle|$ is
of the order of the Bohr radius $a_0$, and $\omega_0$ is of the
order of $E_0/\hbar$, where $E_0=-e^2/8\pi\varepsilon_0a_0$ is the
energy of the ground-state, then $\gamma_0/\omega_0$ is of the order
of $10^{-6}$. For a fixed $z$, the environment induced geometric phase (the second part of Eq.~(\ref{GPb})) reaches its maximum when
$\theta\approx1.354$, which is in the vicinity of $\theta=\pi/2$, i.e., an equal superposition between the ground and excited state. (The numerical results below are based on $\theta=\pi/2$.)

\begin{figure}[htbp]
\centering
\includegraphics[scale=0.8]{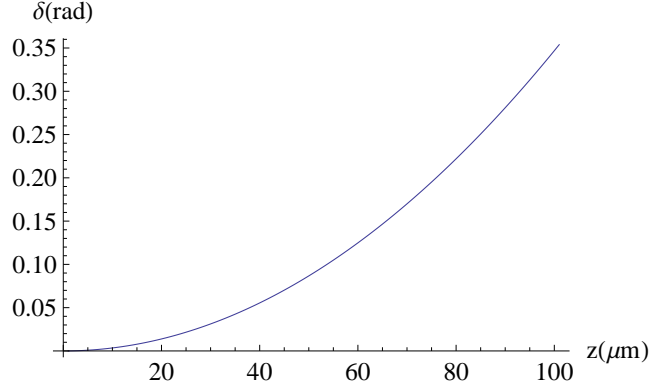}
\caption{ Phase difference $\delta$ as a function of $z$. Here $z$ and $z_0$ are the distances of the trajectories of the two atoms to the plate respectively, with parameters $z_0=1~{\rm \mu m}$, $\omega_0=3\times10^9\,{\rm Hz}$, $\gamma_0/\omega_0=10^{-6}$, and $\theta=\pi/2$.}\label{t1}
\end{figure}

Based upon the discussions above, an experiment that aims a direct
detection of  vacuum fluctuations can be in principle designed. One
first prepares  two-level atoms in a superposition of upper and
lower states in a Ramsey  zone. The atoms are then set in two paths
which are both parallel to a reflecting plane but with different
distances from it.  After a certain time of evolution, we let atoms
from different paths meet by certain means,  for example, by
applying a laser pulse to atoms from one path to change its
direction of motion, and take an interferometric measurement. For a
practical experimental implementation, we want, on one hand, the
difference in the distances of the two paths to the boundary to be
large enough so as to generate an appreciable phase variance. On the
other hand, however, we also want this difference to be negligibly
small as compared to the length of the paths parallel to the plane,
so that the phase accumulated  due to the vertical motion of the
atoms before they meet can be neglected. A compromise can be
achieved for atoms whose transition frequencies are in the microwave
regime, for example, $\omega_0\sim10^{9}~{\rm s}^{-1}$, which is
physically accessible~\cite{freq}. If the distances to the plane
of the parallel paths are chosen respectively as $\sim10^{-5}~{\rm
m}$ and $\sim10^{-6}~{\rm m}$, then one can show by integrating
Eq.~(\ref{bd}) that the geometric phase variation can reach
$\sim10^{-3}~{\rm rad}$ for an evolution time of $\sim10^{-3}~{\rm
s}$. In current cold atom interferometric experiments, the speed of
the atoms is $\sim1~{\rm m/s}$, so the time the atom moves
vertically is $\sim10^{-5}~{\rm s}$, which is two orders of
magnitude less than the parallel evolution time, and thus the phase
accumulated during this period can be neglected. Here we emphasize that the geometric phase is independent of the velocity of the atom. This can be seen from the Fourier transform of the electric-field correlation functions Eqs.~(\ref{fourier})-(\ref{fourier2}), which determine the coefficients of the dissipator $A$ and $B$ (Eq.~(\ref{abc})), and then the geometric phase (Eq.~(\ref{bd})). We specify the  velocity of the atom here only to ensure the geometric phase generated by the motion in the parallel direction will dominate. In Fig.~(\ref{t1}), we plot the phase difference between one atom whose trajectory is fixed at $z_0=1~{\rm \mu m}$, and the other that varies from $z=1~{\rm \mu m}$ to $z=100~{\rm \mu m}$, which shows that the phase difference increases with the distance between the two atoms monotonously. Now, we estimate how the geometric phase would change when the trajectories fluctuate by an amount $\delta z$. We assume one trajectory is fixed at $z_0=1~{\rm \mu m}$, and the other fluctuates from $z=10~{\rm \mu m}$ to $z+\delta z=10.1~{\rm \mu m}$. The phase difference between the  two cases is $\sim10^{-5}~{\rm rad}$, which is two orders of magnitude smaller than $10^{-3}~{\rm rad}$. So the geometric phase is robust against small fluctuations of the distances $\delta z$, as long as $\delta z$ is small compared with $z_0$. Another  effect that should be taken account of is that, in reality, a metal plate does not reflect electromagnetic waves completely. As a result, an excited atom also decays nonradiatively, i.e., the energy is not only transferred to the free space as photons but also to the absorbing metal as heat. The nonradiative decay rate takes the well-known form $\gamma_{non}/\gamma_0=\beta\,z^{-3}$ (see, e.g., Ref.~\cite{absorb1} and references therein, and Ref.~\cite{absorb2} based on a fully canonical quantum theory). This would have an effect on the environment induced geometric phase as can be seen from Eq.~(\ref{GPb}), that is, to the first order, the correction is proportional to the spontaneous emission rate. For conductors, $\beta$ is typically of the order of $\sim10^{-18}~{\rm cm^3}$~\cite{absorb1}, and we are considering $z\sim1~{\rm \mu m}$, so $\gamma_{non}/\gamma_0\sim~10^{-6}$. Therefore, the contribution of the nonradiative decay to the phase can be neglected. We must point out, however, that a subtle issue actually exists  in any practical implementation of our proposal, that is, a cancellation of dynamical phases that the atoms may acquire during the evolution, which is very tricky for systems under non-unitary evolutions like what we are considering here~\cite{remove}. A possible alternative  might be to determine the geometric phase directly  in a tomographic manner by measuring elements of the reduced density matrix of the atom  rather than to perform an interferometric experiment as what is actually pursued  in~\cite{critical} .

\begin{acknowledgments}
This work was supported in part by the National Natural
Science Foundation of China under Grants No. 11075083 and No.
10935013, the Zhejiang Provincial Natural Science Foundation of China under Grant No. Z6100077, the K.C. Wong Magna Fund in Ningbo University, the National Basic Research Program of China under Grant No. 2010CB832803, and the Program for Changjiang Scholars, Innovative Research Team in University (PCSIRT,  No. IRT0964), the Hunan Provincial Natural Science Foundation of China under Grant No. 11JJ7001, and Hunan Provincial Innovation Foundation For Postgraduate under Grant No. CX2012A009.
\end{acknowledgments}


\begin{thebibliography}{00}

\bibitem{lamb}
W. E. Lamb, Jr. and R. C. Retherford, Phys. Rev. {\bf 72}, 241 (1947); H. A. Bethe, Phys. Rev. {\bf 72}, 339 (1947).

\bibitem{casimir1}
H. B. G. Casimir, Proc. K. Ned. Akad. Wet. {\bf 51}, 793 (1948).

\bibitem{casimir2}
H. B. G. Casimir and D. Polder, Phys. Rev. {\bf 73}, 360 (1948).

\bibitem{casimir3}
G. L. Klimchitskaya, U. Mohideen, V. M. Mostepanenko, Rev. Mod. Phys. {\bf 81}, 1827 (2009).

\bibitem{lightcone}
H. Yu and L. H. Ford, Phys. Rev. D {\bf 60}, 084023 (1999); H. Yu and L. H. Ford, Phys. Lett. B {\bf 496}, 107 (2000); H. Yu and P. X. Wu, Phys. Rev. D {\bf 68}, 084019 (2003); H. Yu, N. F. Svaiter and L. H. Ford, Phys. Rev. D {\bf 80}, 124019 (2009).

\bibitem{brown}
H. Yu and L. H. Ford, Phys. Rev. D {\bf 70}, 065009 (2004); H. Yu and J. Chen, Phys. Rev. D {\bf 70} 125006 (2004); H. Yu, X. Fu and P. Wu, J. Phys. A: Math. Theor. {\bf 41}, 335402 (2008); H. Yu, J. Chen and P. Wu, JHEP {\bf 02}, 058 (2006).

\bibitem{cavity}
M. Brune et al., Phys. Rev. Lett. {\bf 72}, 3339 (1994); M. Marrocco, M. Weidinger, R. T. Sang, and H. Walther, Phys. Rev. Lett. {\bf 81}, 5784 (1998).

\bibitem{AHS} A.H. Safavi-Naeini et al,  Phys. Rev. Lett. {\bf 108}, 033602 (2012).

\bibitem{berry}
 M. V. Berry, Proc. R. Soc. Lond. A {\bf 392}, 45 (1984).

\bibitem{GPbook}
{\it Geometric Phases in Physics}, edited by A. Shapere and F.Wilczek (World Scientific, Singapore, 1989).

\bibitem{fault-tolerant}
J. A. Jones, V. Vedral, A. Ekert, and G. Castagnoli, Nature {\bf 403}, 869 (2000).

\bibitem{Uhlmann}
A. Uhlmann, Rep. Math. Phys. {\bf 24}, 229 (1986).

\bibitem{Sjoqvist}
E. Sj\"{o}qvist, A. K. Pati, A. Ekert, J. S.
Anandan, M. Ericsson, D. K. L. Oi, and V. Vedral, Phys. Rev. Lett.
{\bf 85}, 2845 (2000).

\bibitem{Singh}
K. Singh, D. M. Tong, K. Basu, J. L. Chen, and J. F. Du, Phys. Rev. A {\bf 67}, 032106 (2003).

\bibitem{Tong}
D. M. Tong, E. Sj\"{o}qvist, L. C. Kwek, and C. H. Oh, Phys. Rev. Lett. \textbf{93}, 080405 (2004).

\bibitem{Wang}
Z. S. Wang, L. C. Lwek, C. H. Lai, and C. H. Oh,
Europhys. Lett. {\bf 74}, 958 (2006).

\bibitem{Carollo}
A. Carollo, I. Fuentes-Guridi, M. F. Santos, and V. Vedral,
Phys. Rev. Lett. {\bf 90}, 160402 (2003); ibid. {\bf 92}, 020402 (2004).

\bibitem{Rezakhani}
A. T. Rezakhani and P. Zanardi, Phys. Rev. A {\bf 73}, 052117 (2006).

\bibitem{Lombardo}
F. C. Lombardo and P. I. Villar, Phys. Rev. A {\bf 74}, 042311 (2006).

\bibitem{chen}
J. J Chen, J. H. An, Q. J. Tong, H. G. Luo, and C. H. Oh, Phys. Rev. A {\bf 81}, 022120 (2010).

\bibitem{Marzlin}
K.-P. Marzlin, S. Ghose, and B. C. Sanders, Phys. Rev. Lett. {\bf 93}, 260402  (2004).


\bibitem{Du}
J. Du, P. Zou, M. Shi, L. C. Kwek, J. W. Pan, C. H. Oh, A. Ekert, D. K. L. Oi, and M. Ericsson, Phys. Rev. Lett. {\bf 91}, 100403 (2003).

\bibitem{Marie}
M. Ericsson, D. Achilles, J. T. Barreiro, D. Branning, N. A. Peters, and P. G. Kwiat, Phys. Rev. Lett. {\bf 94}, 050401 (2005).

\bibitem{critical}
F. M. Cucchietti, J.-F. Zhang, F. C. Lombardo, P. I. Villar, and R. Laflamme, Phys. Rev. Lett. {\bf 105}, 240406 (2010).

\bibitem{Martin}
E. Martin-Martinez, I. Fuentes and R. B. Mann, Phys. Rev. Lett. {\bf 107}, 131301 (2011).

\bibitem{HuYu}  J. Hu and H. Yu, Phys. Rev. {\bf A 85}, 032105 (2012).

\bibitem{CPP95} G. Compagno, R. Passante, and F. Persico, {\it
Atom-Field Interactions and Dressed Atoms} (Cambridge University Press, Cambridge, England, 1995).

\bibitem{Takagi}
S. Takagi, Prog. Theor. Phys. Suppl. {\bf 88}, 1 (1986).

\bibitem{Borde}
C. J. Bord\'{e}, J. Sharma, P. Tourrenc, and T. Damour, J. Phys. (Paris) {\bf 44}, L983 (1983)

\bibitem{Benatti1}
F. Benatti and  R. Floreanini , Phys. Rev. A {\bf 70}, 012112
(2004).

\bibitem{yu3}
H. Yu and J. Zhang, Phys. Rev. D {\bf 77}, 024031 (2008).


\bibitem{yu4}
H. Yu, Phys. Rev. Lett. {\bf 106}, 061101 (2011).

\bibitem{Lindblad}
V. Gorini, A. Kossakowski, and E. C. G. Surdarshan, J. Math. Phys. {\bf 17}, 821 (1976); G. Lindblad, Commun. Math. Phys. {\bf 48}, 119 (1976).

\bibitem{Benatti2}
F. Benatti and  R. Floreanini,  J. Opt. B {\bf 7}, S429 (2005).

\bibitem{pr5}
F. Benatti, R. Floreanini and M. Piani, Phys. Rev. Lett. {\bf 91}, 070402  (2003).

\bibitem{freq}
J. M. Raimond, M. Brune, and S. Haroche, Rev. Mod. Phys. {\bf 73}, 565 (2001); M. O. Scully, V.V. Kocharovsky, A. Belyanin, E. Fry, and F. Capasso, Phys. Rev. Lett. {\bf 91}, 243004 (2003).

\bibitem{absorb1}
R. R. Chance, A. Prock, and R. Silbey, J. Chem. Phys. {\bf 62}, 2245 (1975).

\bibitem{absorb2}
M. S. Yeung and T. K. Gustafson, Phys. Rev. A {\bf 54}, 5227 (1996).

\bibitem{remove}
E. Sj\"{o}qvist, Physics {\bf 1}, 35 (2008).



\end{thebibliography}
\end{document}